\documentclass[titlecompat,tight,twocolumn,10pt]{article}
\usepackage{usenix}

\usepackage{xspace}
\usepackage{graphicx}
\usepackage{comment}
\usepackage[hyphens]{url}
\let\labelindent\relax
\usepackage{enumitem}
\usepackage{booktabs}
\usepackage{multirow}
\usepackage[binary-units=true]{siunitx}
\usepackage{amsfonts,amsthm,amsmath}
\usepackage{subcaption}
\usepackage{color}
\usepackage[super]{nth}
\usepackage{siunitx}
\usepackage{balance}
\DeclareMathAlphabet{\mathsc}{OT1}{cmr}{m}{sc}
\newtheorem{theorem}{Theorem}[section]

\newtheorem{heuristic}[theorem]{Heuristic}
\newcommand\littleg[1]{}
\newcommand\sarah[1]{}
\newcommand\h[1]{}

\newcommand{\curin}{\mathsf{curIn}}
\newcommand{\curout}{\mathsf{curOut}}
\newcommand{\rate}{\mathsf{rate}}
\newcommand{\userdest}{\mathsf{addr_u}}
\newcommand{\servdest}{\mathsf{addr_s}}
\newcommand{\amt}{\mathsf{amt}}
\newcommand{\tx}{\mathsf{tx}}
\newcommand{\minerfee}{\mathsf{fee}}
\newcommand{\blockbefore}{\delta_b}
\newcommand{\blockafter}{\delta_a}
\newcommand{\status}{\mathsf{status}}
\newcommand{\address}{\mathsf{address}}
\newcommand{\withdraw}{\mathsf{withdraw}}
\newcommand{\incomingCoin}{\mathsf{inCoin}}
\newcommand{\incomingType}{\mathsf{inType}}
\newcommand{\outgoingCoin}{\mathsf{outCoin}}
\newcommand{\outgoingType}{\mathsf{outType}}
\newcommand{\transaction}{\mathsf{tx}}
\newcommand{\transactionURL}{\mathsf{txURL}}
\newcommand{\error}{\mathsf{error}}

\usepackage{hyperref}

\newcommand*\dash{\ifvmode\quitvmode\else\unskip\kern.16667em\fi---%
\hskip.16667em\relax}

\newenvironment{sarahlist}{
\begin{description}[itemsep=2pt,leftmargin=0.4cm]
}{\end{description}}

\begin{document}

\title{Tracing Transactions Across Cryptocurrency Ledgers}

\author{Haaroon Yousaf, George Kappos, and Sarah Meiklejohn\\
	University College London\\
	\texttt{\{h.yousaf,g.kappos,s.meiklejohn\}@ucl.ac.uk}
}

\maketitle

\begin{abstract}
	
	One of the defining features of a cryptocurrency is that its ledger,
	containing all transactions that have ever taken place, 
	is globally visible.  As one consequence of this degree of transparency, a 
	long line of recent research has demonstrated that\dash even in
	cryptocurrencies that are specifically designed to improve anonymity\dash it
	is often possible to track money as it changes hands, and
	in some cases to de-anonymize users entirely.
	With the recent proliferation of alternative cryptocurrencies, however, it
	becomes relevant to ask not only whether or not money can be traced as it
	moves within the ledger of a single cryptocurrency, but if it can in fact be
	traced as it moves \emph{across} ledgers.  This is especially pertinent given
	the rise in popularity of automated trading platforms such as ShapeShift,
	which make it effortless to carry out such cross-currency trades.
	In this paper, we use data scraped from ShapeShift over a thirteen-month period and 
	the data from eight different blockchains to explore this question.
	Beyond developing new heuristics and creating new 
	types of links across cryptocurrency ledgers, we also identify various
	patterns of cross-currency trades and of the general usage of these platforms, 
	with the ultimate goal of understanding whether they serve a criminal 
	or a profit-driven agenda. 
	
\end{abstract}

\section{Introduction}

For the past decade, cryptocurrencies such as Bitcoin have been touted for
their transformative potential, both as a new form of electronic cash and as 
a platform to ``re-decentralize'' aspects of the Internet and computing in
general.  In terms of their role as cash, however, it has been well established 
by now that the usage of pseudonyms in Bitcoin does not achieve meaningful 
levels of
anonymity~\cite{reid2013analysis,FC:RonSha13,FC:AKRSC13,meiklejohn-fistful,FC:SpaMagZan14},
which casts doubt on its role as a payment mechanism.  
Furthermore, the ability to track flows of coins is not limited 
to Bitcoin: it extends even to so-called ``privacy coins'' like
Dash~\cite{FCW:MeiOrl15,malte-2gen},
Monero~\cite{miller2017empirical,ESORICS:KFTS17,HH19,yu-monero}, and
Zcash~\cite{zcash-anon,sarah-zcash} that incorporate features explicitly 
designed to improve on Bitcoin's anonymity guarantees.

Traditionally, criminals attempting to cash out illicit funds would have
to use exchanges;
indeed, most tracking techniques rely on identifying the addresses associated
with these exchanges as a way to observe when these deposits
happen~\cite{meiklejohn-fistful}.  Nowadays, however, exchanges 
typically implement strict Know Your Customer/Anti-Money Laundering 
(KYC/AML) policies to comply with regulatory requirements, meaning 
criminals (and indeed all users) risk revealing their real identities when 
using them.  Users also run risks when storing their coins in accounts at 
custodial 
exchanges, as exchanges may be hacked or their coins may otherwise become
inaccessible~\cite{mtgoxwired,quadriga}.  As an alternative, there
have emerged in the past few years frictionless trading platforms such 
as ShapeShift\footnote{\url{https://shapeshift.io}} and
Changelly,\footnote{\url{https://changelly.com}} in which users are 
able to trade between cryptocurrencies without having to store their coins
with the platform provider.  Furthermore, while ShapeShift now requires users
to have verified accounts~\cite{ss-id}, this was not the case before October
2018.

Part of the reason for these trading platforms to exist is the sheer rise in
the number of different cryptocurrencies: according to the popular
cryptocurrency data tracker CoinMarketCap there were 36
cryptocurrencies in September 2013, only 7 of which had a stated market
capitalization of over 1 million
USD,\footnote{\url{https://coinmarketcap.com/historical/20130721/}} whereas
in January 2019 there were 2117 cryptocurrencies, of which the top 10 had a market
capitalization of over 100 million USD.  
Given this proliferation of new cryptocurrencies and platforms that make it
easy to transact across them, it becomes important to 
consider not just whether or not flows of coins can be tracked within the
transaction ledger of a given currency, but also if they can be tracked as
coins move across their respective ledgers as well.  This 
is especially important given that there are documented cases of criminals
attempting to use these cross-currency trades to obscure the flow of their 
coins: the WannaCry ransomware operators, for example, 
were observed using ShapeShift to convert their ransomed bitcoins into
Monero~\cite{wannacry}.  More generally, these services have the potential to 
offer an insight into the broader cryptocurrency ecosystem and the thousands
of currencies it now contains.

In this paper, we initiate an exploration of the usage of these 
cross-currency trading platforms, and the potential 
they offer in terms of the ability to track flows 
of coins as they move across different transaction ledgers.  Here we rely on 
three distinct sources of data: the cryptocurrency blockchains, 
the data collected via our own interactions with these trading platforms, 
and\dash as we describe in Section~\ref{sec:data}\dash the information offered 
by the platforms themselves via their public APIs. 

We begin in Section~\ref{sec:phases} by identifying the specific on-chain 
transactions associated with an advertised ShapeShift transaction, which we 
are able to do with a relatively high degree of success (identifying both the 
deposit and withdrawal transactions 81.91\% of the time, on average).  We 
then describe in
Section~\ref{sec:tracking} the different transactional patterns that can be 
traced by identifying the relevant on-chain transactions, focusing
specifically on patterns that may be indicative of trading or money
laundering, and on the ability to link addresses across different currency
ledgers.  We then move in
Section~\ref{sec:clusters} to consider both old and new heuristics for
clustering together addresses associated with ShapeShift, with particular
attention paid to our new heuristic concerning the common social relationships
revealed by the usage of ShapeShift.  Finally, we bring all the analysis
together by applying it to several case studies in
Section~\ref{sec:case-studies}.  Again, our particular focus in this last 
section is on the phenomenon of trading and other profit-driven activity, 
and the extent to which usage of the 
ShapeShift platform seems to be motivated by criminal activity or a more 
general desire for anonymity.

\section{Related Work}\label{sec:related}

We are not aware of any other research exploring these cross-currency trading
platforms, but consider as related all research that explores the level of 
anonymity achieved by cryptocurrencies.  This work is complementary to 
our own, as the techniques it develops can be combined with ours to track the 
entire flow of cryptocurrencies as they move both within and across different 
ledgers.

Much of the earlier research in this vein focused on
Bitcoin~\cite{reid2013analysis,FC:RonSha13,FC:AKRSC13,meiklejohn-fistful,FC:SpaMagZan14},
and operates by adopting the so-called ``multi-input'' heuristic, which says
that all input addresses in a transaction belong to the same entity (be it an
individual or a service such as an exchange).  While the accuracy of this
heuristic has been somewhat eroded by privacy-enhancing techniques like
CoinJoin~\cite{maxwell2013coinjoin}, new techniques have been developed to avoid such
false positives~\cite{malte-2gen}, and as such it has now been 
accepted as standard and incorporated into many tools for Bitcoin blockchain 
analytics.\footnote{\url{https://www.chainalysis.com/}}\footnote{\url{https://www.elliptic.co/}}  
Once addresses are clustered together in this manner,
the entity can then further be identified using hand-collected tags that form
a ground-truth dataset.  We adopt both of these techniques in order to analyze
the clusters formed by ShapeShift and Changelly in a variety of cryptocurrency
blockchains, although as described in Section~\ref{sec:clusters} we find them
to be relatively unsuccessful in this setting.

In response to the rise of newer ``privacy coins'', a recent line of research
has also worked to demonstrate that the deployed versions of these
cryptocurrencies have various properties that diminish the level of anonymity 
they achieve in practice.  This includes work targeting
Dash~\cite{malte-2gen,FCW:MeiOrl15},
Monero~\cite{miller2017empirical,ESORICS:KFTS17,HH19,yu-monero}, and
Zcash~\cite{zcash-anon,sarah-zcash}.  

In terms of Dash, its main privacy
feature is similar to CoinJoin, in which different senders join forces to
create a single transaction representing their transfer to a diverse set of
recipients.  Despite the intention for this to hide which recipient addresses
belong to which senders, research has demonstrated that such links can in 
fact be created based on the value being 
transacted~\cite{malte-2gen,FCW:MeiOrl15}.  
Monero, which allows senders to hide which input belongs to them by using
``mix-ins'' consisting of the keys of other users, is vulnerable to
de-anonymization attacks exploiting the (now-obsolete) case in which some 
users chose not to use mix-ins, or exploiting inferences about the age of the
coins used as mix-ins~\cite{miller2017empirical,ESORICS:KFTS17,HH19,yu-monero}.
Finally, Zcash is similar to Bitcoin, but with the addition of a privacy
feature called the shielded pool, which can be used to hide the values and
addresses of the senders and recipients involved in a transaction. Recent
research has shown that it is possible to significantly reduce the anonymity 
set provided by the shielded pool, by developing simple heuristics for 
identifying links between hidden and partly obscured
transactions~\cite{zcash-anon,sarah-zcash}.

\section{Background}\label{sec:back}

\subsection{Cryptocurrencies}\label{sec:back-currency}

The first decentralized cryptocurrency, Bitcoin, was
created by Satoshi Nakamoto in 2008~\cite{satoshi-bitcoin} and deployed in
January 2009.  At the most basic level, bitcoins are digital assets that can 
be traded between sets of users without the need for any trusted intermediary.
Bitcoins can be thought of as being stored in a public key, which is
controlled by the entity in possession of the associated private key.
A single user can store their assets across many public keys, which act
as pseudonyms with no inherent link to the user's identity.  In order to spend
them, a user can form and cryptographically sign a transaction that acts to
send the bitcoins to a recipient of their choice.  
Beyond Bitcoin, other platforms now offer more robust functionality.  For
example, Ethereum allows users to deploy \emph{smart contracts} onto the
blockchain, which act as stateful programs that can be triggered by
transactions providing inputs to their functions.  

In order to prevent double-spending, many cryptocurrencies are 
\emph{UTXO-based}, meaning coins are
associated not with an address but with a uniquely identifiable UTXO (unspent
transaction output) that is created for all outputs in a given transaction.
This means that one address could be associated with potentially many
UTXOs (corresponding to each time it has received coins), and that inputs 
to transactions are also UTXOs rather than addresses.  
Checking for double-spending is then just a matter of checking
if an input is in the current UTXO`set, and removing it from the set once it
spends it contents.

\subsection{Digital asset trading platforms}\label{sec:back-shapeshift}

In contrast to a traditional (custodial) exchange, a digital asset trading 
platform allows users to move between different cryptocurrencies without
storing any money in an account with the service; in other words,
users keep their own money in their own accounts and the platform has it only
at the time that a trade is being executed.  To initiate such a trade, a user 
approaches the service and selects a supported input currency $\curin$ (i.e., 
the currency from which they would like to move money) and a supported output 
currency $\curout$ (the currency that they would like to obtain).  
A user additionally specifies a destination
address $\userdest$ in the $\curout$ blockchain, which is the address to which 
the output currency will be sent.  The service then presents the user with 
an exchange rate $\rate$ and an address $\servdest$ in the $\curin$ blockchain 
to which to send money, as well as a miner fee $\minerfee$ that accounts for
the transaction it must form in the $\curout$ blockchain.  The user then 
sends to this address $\servdest$ the amount $\amt$ in 
$\curin$ they wish to convert, and after some delay the service sends the 
appropriate amount of the output currency to the specified destination 
address $\userdest$.  This means that an interaction with these services
results in two transactions: one on the $\curin$ blockchain sending $\amt$ to
$\servdest$, and one on the $\curout$ blockchain sending (roughly)
$\rate\cdot\amt - \minerfee$ to $\userdest$.

This describes an interaction with an abstracted platform.  Today, the two 
best-known examples are ShapeShift and Changelly.  Whereas Changelly has
always required account creation, ShapeShift introduced this requirement only
in October 2018.  Each platform supports
dozens of cryptocurrencies, ranging from better-known ones such as Bitcoin and
Ethereum to lesser-known ones such as FirstBlood and Clams.  
In Section~\ref{sec:data}, we describe in more depth the operations of these 
specific platforms and our own interactions with them.

\section{Data Collection and Statistics}\label{sec:data}

In this section, we describe our data sources, as well as some preliminary
statistics about the collected data.  We begin in
Section~\ref{sec:changelly-data} by describing our own 
interactions with Changelly, a trading platform with a
limited personal API.  We then describe in Section~\ref{sec:shapeshift-data} 
both our own interactions with
ShapeShift, and the data we were able to scrape from their public API, which
provided us with significant insight into their overall set of transactions.
Finally, we describe in Section~\ref{sec:blockchain-data} our collection of
the data backing eight different cryptocurrencies.

\subsection{Changelly}\label{sec:changelly-data}

Changelly offers a simple API\footnote{\url{https://api-docs.changelly.com/}}
that allows registered users to carry out transactions with
the service.  Using this API, we engaged in 22 transactions, using the most
popular ShapeShift currencies (Table~\ref{tab:popular}) to guide our choices 
for $\curin$ and $\curout$. 

While doing these transactions, we observed that they would sometimes take 
up to an hour to complete.  This is because Changelly attempts to minimize 
double-spending risk by requiring users to wait for a set number 
of confirmations (shown to the user at the time of their transaction) in the 
$\curin$ blockchain before executing the transfer on the $\curout$ blockchain.  
We used this observation to
guide our choice of parameters in our identification of on-chain transactions
in Section~\ref{sec:phases}.

\subsection{ShapeShift}\label{sec:shapeshift-data}

ShapeShift's API\footnote{\url{https://info.shapeshift.io/api}} allows 
users to execute their own transactions, of which we did 18 in total.  %
As with Changelly, we were able to gain some valuable insights about the
operation of the platform via these personal interactions.  
Whereas ShapeShift did not disclose the number of confirmations they
waited for on the $\curin$ blockchain, we again observed long delays,
indicating that they were also waiting for a sufficient number.

Beyond these personal interactions, the API provides information on 
the operation of the service as 
a whole. Most notably, it provides three separate pieces of information: (1)
the current trading rate between any pair of cryptocurrencies, (2) 
a list of up to 50 of the most recent transactions that have taken place 
(across all users), and (3) full details of a specific ShapeShift transaction 
given the address $\servdest$ in the $\curin$ blockchain (i.e., the address to
which the user sent their coins).

For the trading rates, ShapeShift provides the following information for all
cryptocurrency pairs $(\curin,\curout)$: the rate, the limit (i.e., the 
maximum that can be exchanged), the minimum that can be exchanged, and the 
miner fee (denominated in $\curout$).  For the 50 most recent transactions, 
information is provided in the form:
$(\curin,\curout,\amt,t,\mathsf{id})$, where the first three of these are as
discussed in Section~\ref{sec:back-shapeshift}, $t$ is a UNIX timestamp, and
$\mathsf{id}$ is an internal identifier for this transaction.  For the
transaction information, when provided with a specific $\servdest$ ShapeShift
provides the tuple $(\status,\address,\withdraw,\incomingCoin,
\incomingType,\outgoingCoin,\allowbreak \outgoingType,\transaction,\transactionURL,\error)$.
The $\status$ field is a flag that is either \verb#complete#, to mean the
transaction was successful; \verb#error#, to mean an issue occurred with the
transaction or the queried address was not a ShapeShift address; or
\verb#no_deposits#, to mean a user initiated a transaction but did not send
any coins.  The $\error$ field appears when an error is returned and gives a
reason for the error.  The $\address$ field is the same address $\servdest$
used by ShapeShift, and $\withdraw$ is the address $\userdest$ (i.e., the
user's recipient address in the $\curout$ blockchain).  $\incomingType$ and
$\outgoingType$ are the respective $\curin$ and $\curout$ currencies and 
$\incomingCoin$ is the $\amt$ received.  $\outgoingCoin$ is the amount sent in
the $\curout$ blockchain.  Finally, $\transaction$ is the transaction hash in
the $\curout$ blockchain and $\transactionURL$ is a link to this transaction 
in an online explorer. 

Using a simple Web scraper, we downloaded the transactions and rates every
five seconds for close to thirteen months: from November 27 
2017 until December 23 2018.  This resulted in a set of 
2,843,238 distinct transactions. Interestingly, we noticed that several 
earlier test transactions we did with the platform did
not show up in their list of recent transactions, which suggests that their
published transactions may in fact underestimate their overall activity.

\subsubsection{ShapeShift currencies}

\begin{figure}
	\centering
	\includegraphics[width=\linewidth]{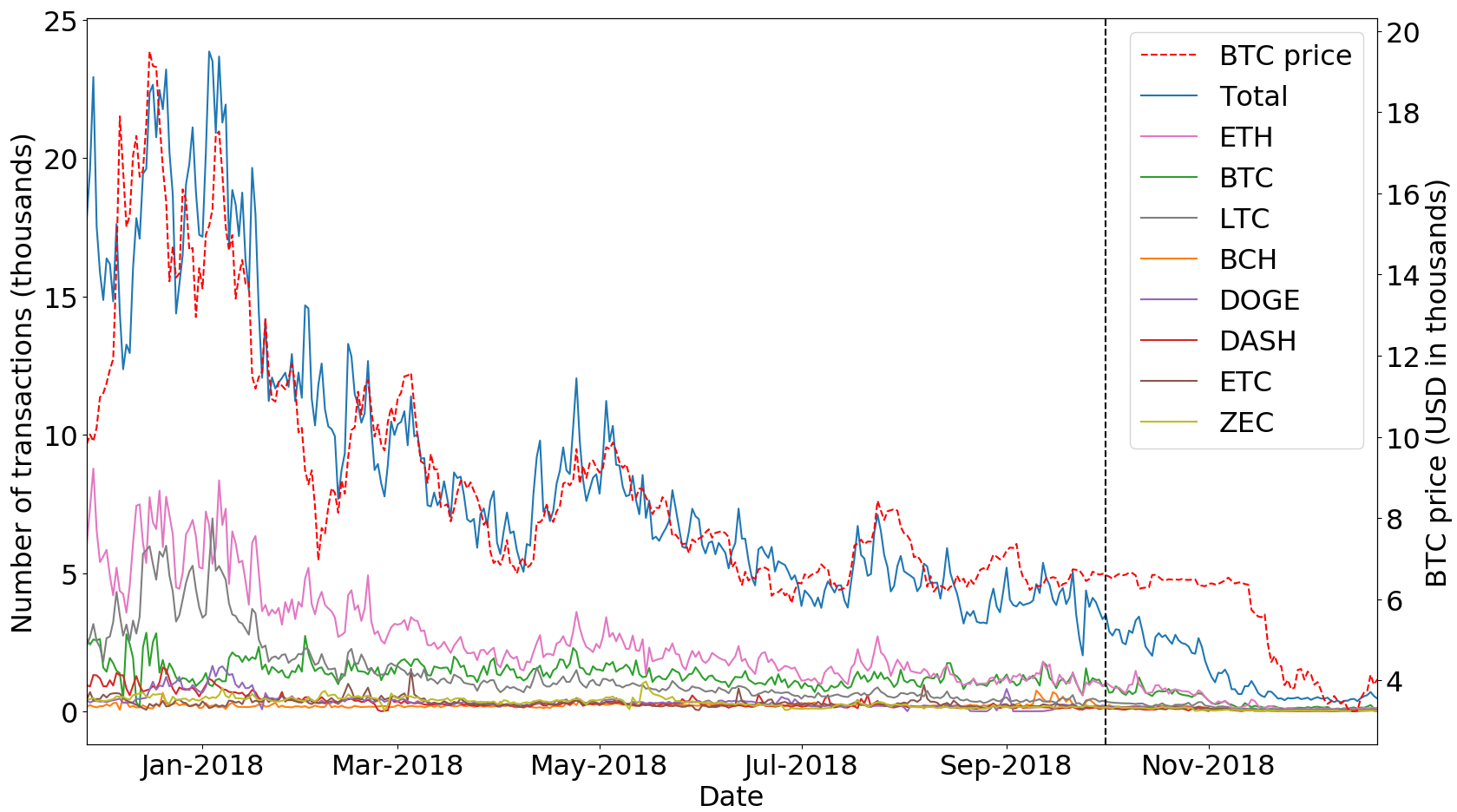}
	\caption{The total number of transactions per day reported via ShapeShift's API,
		and the numbers broken down by cryptocurrency (where a transaction is attributed 
		to a coin if it is used as either $\curin$ or $\curout$).  The dotted red line
		indicates the BTC-USD exchange rate, and the horizontal dotted black line
		indicates when KYC was introduced into ShapeShift.}
	\label{fig:ss-time-distro}
\end{figure}

In terms of the different cryptocurrencies used in ShapeShift transactions, 
their popularity was distributed as seen in
Figure~\ref{fig:ss-time-distro}.  As this figure depicts, the overall
activity of ShapeShift is (perhaps unsurprisingly) correlated with the price
of Bitcoin in the same time period.  At the same time, there is a 
decline in the number of transactions after KYC was introduced that is not
clearly correlated with the price of Bitcoin (which is largely steady and
declines only several months later).

ShapeShift supports dozens of cryptocurrencies, and in our data we observed
the use of 65 different ones.   The most commonly
used coins are shown in Table~\ref{tab:popular}.  %
It is clear that Bitcoin and Ethereum are 
the most heavily used currencies, which is perhaps not surprising given the 
relative ease with which they can be exchanged with fiat currencies on more 
traditional exchanges, and their rank in terms of market capitalization.  

\begin{table}[t]
	\centering
	{\footnotesize
		\begin{tabular}{lcS[table-format=7.0]S[table-format=6.0]S[table-format=6.0]}
			\toprule
			Currency & Abbr. & Total & $\curin$ & $\curout$ \\
			\midrule
			Ethereum & ETH & 1385509 & 892971 & 492538 \\
			Bitcoin  & BTC & 1286772 & 456703 & 830069 \\
			Litecoin & LTC & 720047 & 459042 & 261005 \\
			Bitcoin Cash & BCH & 284514 & 75774 & 208740 \\
			Dogecoin & DOGE & 245255 & 119532 & 125723 \\
			Dash & DASH & 187869 & 113272 & 74597 \\
			Ethereum Classic & ETC & 179998 & 103177 & 76821 \\
			Zcash & ZEC & 154142 & 111041 & 43101 \\
			\bottomrule
		\end{tabular}
	}
	\caption{The eight most popular coins used on ShapeShift, in terms of the total 
		units traded, and the respective units traded with that coin as $\curin$ and 
		$\curout$.}
	\label{tab:popular}
\end{table}

\subsection{Blockchain data}\label{sec:blockchain-data}

For the cryptocurrencies we were interested in exploring further, it was also
necessary to download and parse the respective blockchains, in order to 
identify the on-chain transactional behavior of ShapeShift and Changelly.  
It was not feasible to do this for all 65 currencies used on ShapeShift
(not to mention that given the low volume of transactions for many of them, 
it would likely not yield additional insights anyway), so we chose to focus 
instead on just the top 8, as seen in Table~\ref{tab:popular}.  
Together, these 
account for 95.7\% of all ShapeShift transactions if only one of $\curin$ 
and/or $\curout$ is one of the eight, and 60.5\% if both are.  

For each of these currencies, we ran a full node in order to download the
entire blockchain.  For the ones supported by the BlockSci tool~\cite{blocksci} 
(Bitcoin, Dash and Zcash), 
we used it to parse and analyze their blockchains.  BlockSci does not,
however, support the remaining five currencies.  For these we
thus parsed the blockchains using Python scripts, stored the data as Apache 
Spark parquet files, and analyzed them using custom scripts.  In total, we
ended up working with 654~GB of raw blockchain data and 434~GB of 
parsed blockchain data.

\section{Identifying Blockchain Transactions}\label{sec:phases}

In order to gain deeper 
insights about the way these trading platforms are used, it is necessary to 
identify 
not just their internal transactions but also the transactions that appear on
the blockchains of the traded currencies.  This section presents heuristics
for identifying these on-chain transactions, and the next section explores
the additional insights these transactions can offer.

Recall from Section~\ref{sec:back-shapeshift} that an interaction with
ShapeShift
results in the deposit of coins from the user to the service on the $\curin$ 
blockchain (which we refer to as ``Phase~1''), and the withdrawal of coins
from the service to the user on the $\curout$ blockchain
(``Phase~2'').  
To start with Phase~1, we thus seek to identify the deposit transaction on 
the input ($\curin$) blockchain.  Similarly to Portnoff et
al.~\cite{portnoff-kdd}, we consider two main requirements for
identifying the correct on-chain transaction: (1) that it 
occurred reasonably close in time to the point at which it was
advertised via the API, and (2) that the value it carried was identical to the 
advertised amount.

For this first requirement, we look for candidate transactions as follows.  
Given a ShapeShift transaction with timestamp $t$, we first find the block 
$b$ (at some height 
$h$) on the $\curin$ blockchain that was mined at the time closest to $t$.  We 
then look at the transactions in all blocks with height in the range 
$[h-\blockbefore,h+\blockafter]$, where $\blockbefore$ and $\blockafter$ 
are parameters specific to $\curin$.  We looked at both earlier and
later blocks based on the observation in our own interactions that the
timestamp published by ShapeShift would sometimes be earlier and sometimes be
later than the on-chain transaction.

For each of our eight currencies, we ran this heuristic for every ShapeShift 
transaction using $\curin$ as the currency in question, with every possible
combination of $\blockbefore$ and $\blockafter$ ranging from $0$ to $30$.
This resulted in a set of candidate transactions with zero hits (meaning no
matching transactions were found), a single hit, or multiple hits.  To rule out
false positives, we initially considered as successful only ShapeShift 
transactions with a single candidate on-chain transaction, although we
describe below an augmented heuristic that is able to tolerate multiple hits.  
We then used the values of $\blockbefore$ and $\blockafter$ that maximized 
the number of single-hit transactions for each currency.  
As seen in Table~\ref{tab:phase12}, the optimal choice of these parameters 
varies significantly across currencies, according to their different block 
rates; typically we needed to look further before or after for currencies in
which blocks were produced more frequently.  

In order to validate the results
of our heuristic for Phase~1, we use the additional capability of the ShapeShift 
API described in Section~\ref{sec:shapeshift-data}. In particular, we queried the 
API on the recipient address of every transaction identified by our heuristic 
for Phase~1.  If the response of the API was affirmative, we flagged the recipient 
address as belonging to ShapeShift and we identified the transaction in which it
received coins as the $\curin$ transaction.  This also provided a way to
identify the corresponding Phase~2 transaction on the $\curout$ blockchain,
as it is just the $\transaction$ field returned by the API.  As we
proceed only in the case that the API returns a valid result, we gain
ground-truth data in both Phase~1 and Phase~2.  In other words, this method
serves to not only validate our results in Phase~1 but also provides a way to
identify Phase~2 transactions.

The heuristic described above is able to handle only single hits; i.e., the
case in which there is only a single candidate transaction.  Luckily, it is
easy to augment this heuristic by again using the API.  For example, assume we 
examine
a BTC-ETH ShapeShift transaction and we find three candidate transactions in
the Bitcoin blockchain after applying the basic heuristic described above.  To 
identify which of
these transactions is the right one, we simply query the API on all three
recipient addresses and check that the $\status$ field is affirmative (meaning
ShapeShift recognizes this address) and that the $\outgoingType$ field is ETH.
In the vast majority of cases this uniquely identifies the correct transaction
out of the candidate set, meaning we can use the API to both validate our
results (i.e., we use it to eliminate potential false positives, as described
above) and to augment the heuristic by being able to tolerate multiple
candidate transactions.  The augmented results for Phase~1 can be found in the
last column of Table~\ref{tab:phase12} and clearly demonstrate the benefit
of this extra usage of the API.  In the most dramatic example, we were able to
go from identifying the on-chain transactions for ShapeShift transactions
involving Bitcoin 65.75\% of the time with the basic heuristic to
identifying them 76.86\% of the time with the augmented heuristic.

\begin{table}[t]
	\centering
	\begin{tabular}{lccS[table-format=2.2]S[table-format=2.2]}
		\toprule
		Currency & \multicolumn{2}{c}{Parameters} & {Basic \%} & {Augmented \%} \\
		\cmidrule(lr){2-3}
		& {$\blockbefore$} & {$\blockafter$} && \\
		\midrule
		BTC & 0 & 1 & 65.76 & 76.86 \\
		BCH & 9 & 4 & 76.96 & 80.23 \\
		DASH & 5 & 5 & 84.77 & 88.65  \\
		DOGE & 1 & 4 & 76.94 & 81.69  \\
		ETH & 5 & 0 & 72.15 & 81.63 \\
		ETC & 5 & 0 & 76.61 & 78.67  \\
		LTC & 1 & 2 & 71.61 & 76.97  \\
		ZEC & 1 & 3 & 86.94 & 90.54 \\
		\bottomrule
	\end{tabular}
	\caption{For the selected (optimal) parameters and for a given currency used
		as $\curin$, the percentage of ShapeShift transactions for which we found 
		matching on-chain transactions for both the basic (time- and value-based) 
		and the augmented (API-based) Phase~1
		heuristic.  The augmented heuristic uses the API and thus also represents 
		our success in identifying Phase~2 transactions.}
	\label{tab:phase12}
\end{table}

\subsection{Accuracy of our heuristics}

False negatives can occur for both of our heuristics when there are either too 
many or too few matching transactions in the searched block interval.  These 
are more common for the basic heuristic, as described above and seen in 
Table~\ref{tab:phase12}, because it is conservative in identifying an
on-chain transaction only when there is one candidate.  This rate could be
improved by increasing the searched block radius, at the expense of adding
more computation and potentially increasing the false positive rate.

False positives can occur for both of our heuristics if someone sends the same 
amount as the ShapeShift transaction at roughly the same time, but this 
transaction falls within our searched interval whereas the ShapeShift one 
doesn't.  In theory, this
should not be an issue for our augmented heuristic, since the API
will make it clear that the candidate transaction is not in fact associated
with ShapeShift.  In a small number of cases (fewer than 1\% of all ShapeShift 
transactions), however, the API returned details of a transaction with different
characteristics than the one we were attempting to identify; e.g., it had a
different pair of currencies or a different value being sent.  This happened 
because ShapeShift allows users to re-use an existing deposit address, and the 
API returns only the latest transaction using a given address. 

If we blindly took the results of the API, then this would lead to false 
positives in our augmented heuristic for both Phase~1 and Phase~2.  We thus 
ensured that the transaction returned by the
API had three things in common with the ShapeShift transaction:
(1) the pair of currencies, (2) the amount being sent, and (3) the timing,
within the interval specified in Table~\ref{tab:phase12}.  If there was any 
mismatch, we discarded the transaction.  For example, given a ShapeShift
transaction indicating an ETH-BTC shift carrying 1~ETH and occurring at 
time $t$, we looked for all addresses that received 1~ETH at time $t$ or up to 
$5$ blocks earlier.  We then queried the API on these addresses and kept only
those transactions which reported shifting 1~ETH to BTC.  
While our augmented heuristic still might produce false positives in the
case that a user quickly makes two different transactions using the same 
currency pair, value, and deposit address, we view this as unlikely, 
especially given the relatively long wait times we observed ourselves when
using the service (as mentioned in Section~\ref{sec:shapeshift-data}).

\subsection{Alternative Phase~2 identification}

Given that our heuristic for Phase~2 involved just querying the API for the
corresponding Phase~1 transaction, it is natural to wonder what would be
possible without this feature of the API, or indeed if there are any
alternative strategies for identifying Phase~2 transactions.  Indeed, it is
possible to use a similar heuristic for identifying Phase~1 transactions, by
first looking for transactions in blocks that were mined close to the 
advertised transaction time, and then looking for ones in which the amount was
close to the expected amount.  Here the amount must be estimated according to
the advertised $\amt$, $\rate$, and $\minerfee$.  In theory, the amount sent 
should be $\amt\cdot\rate - \minerfee$, although in practice the rate can
fluctuate so it is important to look for transactions carrying a total value 
within a reasonable error rate of this amount.

When we implemented and applied this heuristic, we found that our accuracy 
in identifying Phase~2 transactions decreased
significantly, due to the larger set of transactions that carried an amount
within a wider range (as opposed to an exact amount, as in Phase~1) and the
inability of this type of heuristic to handle multiple candidate transactions.
More importantly, this approach provides no ground-truth information at all:
by choosing conservative parameters it is possible to limit the number of
false positives, but this is at the expense of the false negative rate (as,
again, we observed in our own application of this heuristic) and in general
it is not guaranteed that the final set of transactions really are associated
with ShapeShift.  As this is the exact guarantee we can get by using the API,
we continue in the rest of the paper with the results we obtained there, but
nevertheless mention this alternative approach in case this feature of the API
is discontinued or otherwise made unavailable. 

\section{Tracking Cross-Currency Activity}\label{sec:tracking}

In the previous section, we saw that it was possible in many cases to identify
the on-chain transactions, in both the $\curin$ and $\curout$ blockchains, 
associated with the transactions advertised by
ShapeShift.  In this section, we take this a step further and show how
linking these transactions can be used to identify more complex patterns 
of behavior.

\begin{figure*}[t]
	\centering
	\begin{subfigure}[b]{0.34\textwidth}
		\centering
		\includegraphics[width=\linewidth]{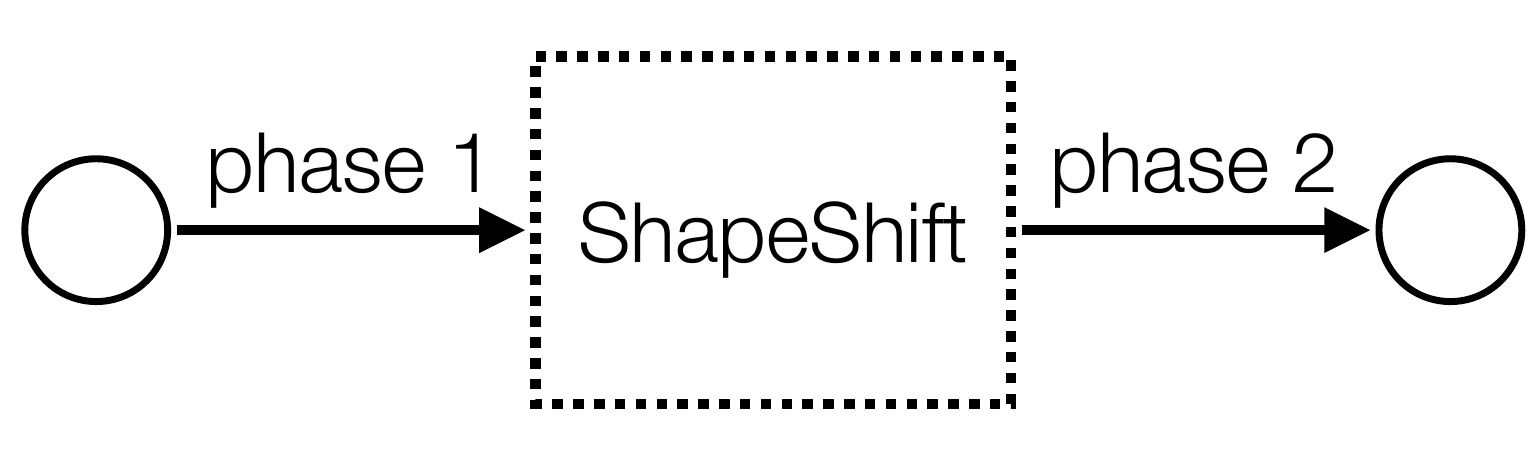}
		\caption{Pass-through}
		\label{fig:passthrough}
	\end{subfigure}
	~
	\begin{subfigure}[b]{0.24\textwidth}
		\centering
		\includegraphics[width=\linewidth]{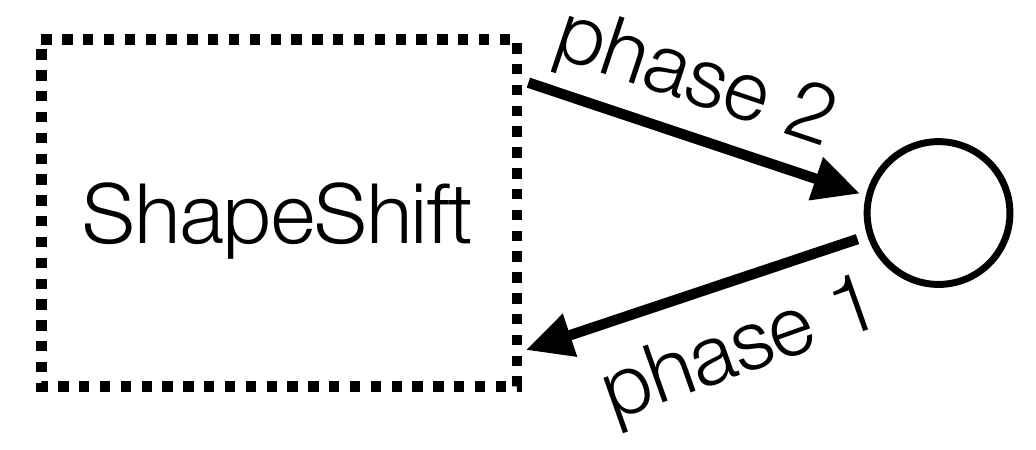}
		\caption{U-turn}
		\label{fig:uturn}
	\end{subfigure}
	~
	\begin{subfigure}[b]{0.34\textwidth}
		\centering
		\includegraphics[width=\linewidth]{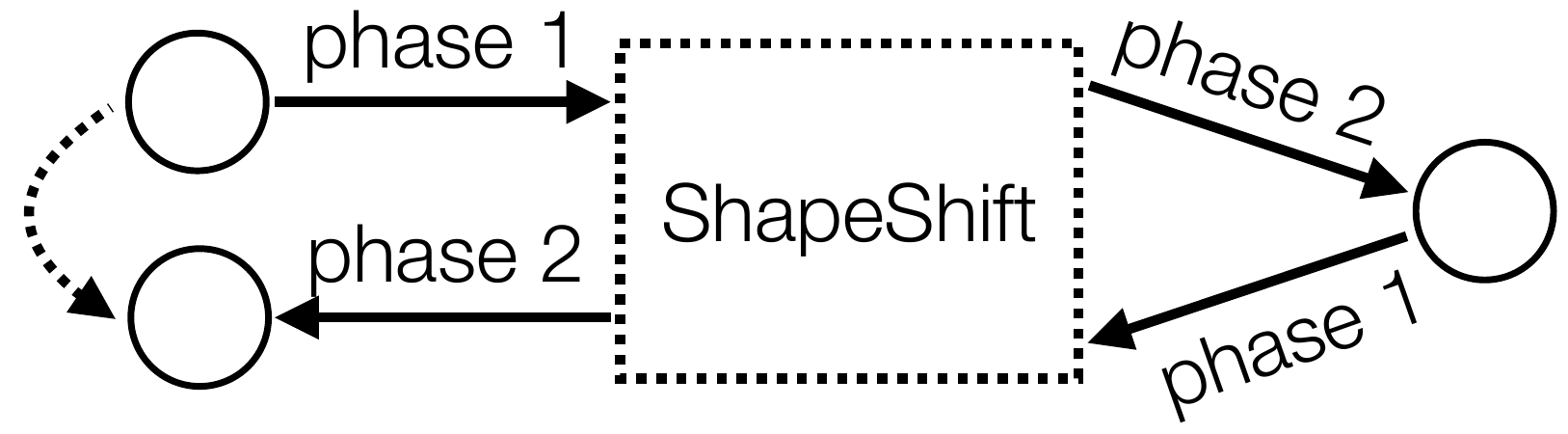}
		\caption{Round-trip}
		\label{fig:roundtrip}
	\end{subfigure}
	\caption{The different transactional patterns, according to how
		they interact with ShapeShift and which phases are required to identify them.}
	\label{fig:tx-types}
\end{figure*}

As shown in Figure~\ref{fig:tx-types}, we consider these for three main types
of transactions.  In particular, we look at (1) \emph{pass-through}
transactions, which represent the full flow of money as it moves from one
currency to the other via the deposit and withdrawal transactions; 
(2) \emph{U-turns}, in which a user who has shifted into one currency 
immediately shifts back; and (3) \emph{round-trip} transactions, which are
essentially a combination of the first two and follow a user's flow of money
as it moves from one currency to another and then back to the original one.
Our interest in these particular patterns of behavior is largely based on the 
role they play in tracking money as it moves across the ledgers of different 
cryptocurrencies.  In particular, our goal is to test the validity of the
implicit assumption made by criminal usage of the platform\dash such as we
examine further in Section~\ref{sec:case-studies}\dash that ShapeShift 
provides additional anonymity beyond simply transacting in a given currency.

In more detail, identifying pass-through transactions allows us to create a 
link between the input address(es) in the deposit on the $\curin$ 
blockchain and the output address(es) in the withdrawal on the 
$\curout$ blockchain.

Identifying U-turns allows us to see when a user has interacted with
ShapeShift not because they are interested in holding units of the $\curout$
cryptocurrency, but because they see other benefits in shifting coins back and
forth.  There are several possible motivations for this: for example, traders 
may quickly shift back and forth between two different cryptocurrencies in 
order to profit from differences in their price.  We investigate this
possibility in Section~\ref{sec:trading-bots}.  Similarly, people performing 
money laundering or 
otherwise holding ``dirty'' money may engage in such behavior under the belief 
that once the coins are moved back into the $\curin$ blockchain, they are 
``clean'' after moving through ShapeShift regardless of what happened with the 
coins in the $\curout$ blockchain.

Finally, identifying round-trip transactions allows us to create a link
between the input address(es) in the deposit on the $\curin$ 
blockchain with the output address(es) in the later withdrawal on 
the $\curin$ blockchain.  %
Again, there are many reasons why users might engage in such
behavior, including the trading and money laundering examples given above.  As
another example, if a $\curin$ user wanted to make an anonymous payment to 
another $\curin$ user, they might attempt to do so via a round-trip
transaction (using the address of the other user in the second pass-through
transaction), under the same assumption that ShapeShift would sever the link
between their two addresses.

\subsection{Pass-through transactions}\label{sec:pass}

\begin{figure}[t]
	\centering
	\includegraphics[width=0.7\linewidth]{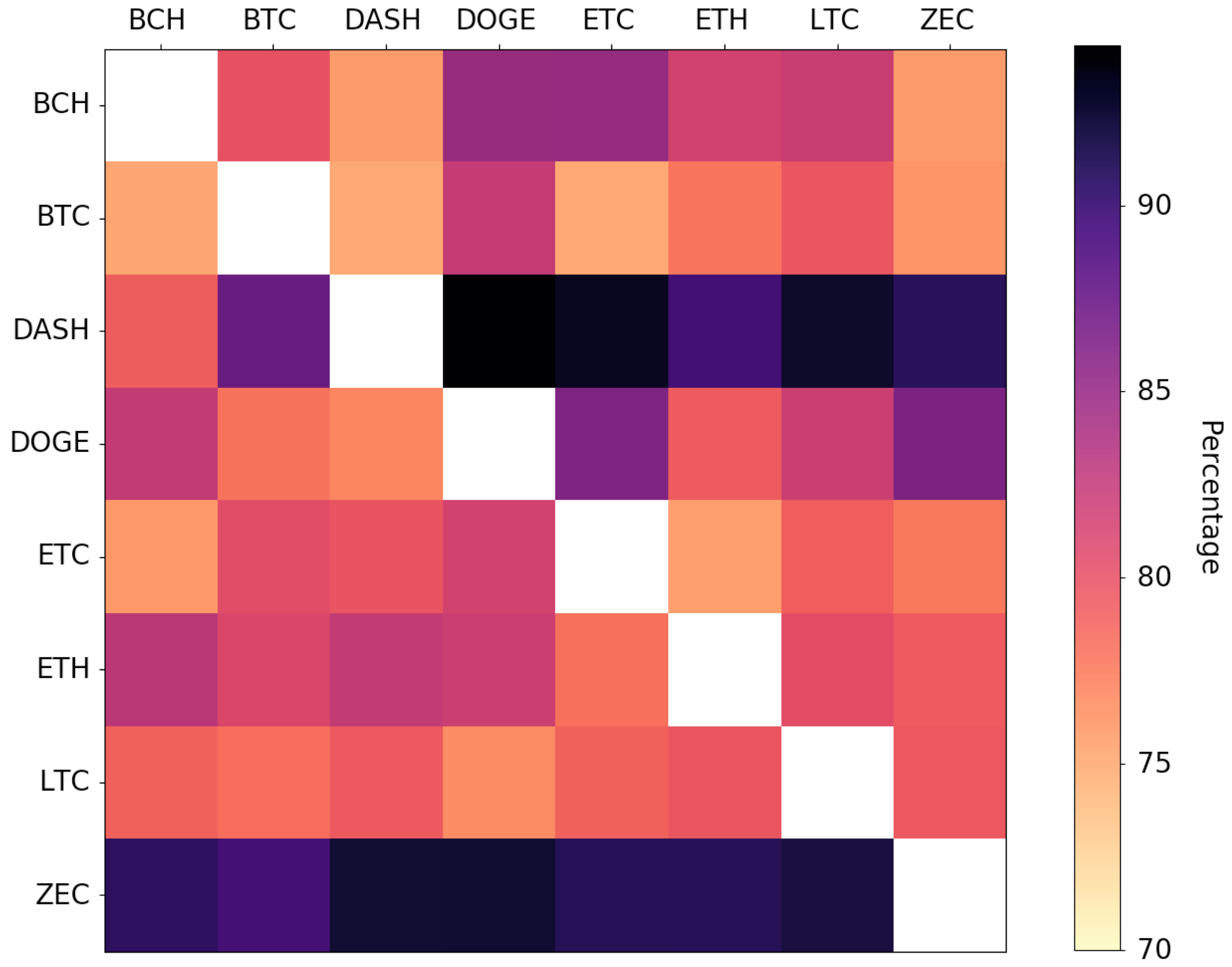}
	\caption{For each pair of currencies, the number of transactions we identified 
		as being a pass-through from one to the other, as a percentage of 
		the total number of transactions between those two currencies.}
	\label{fig:passthrough-heatmap}
\end{figure}

Given a ShapeShift transaction from $\curin$ to $\curout$, the methods from
Section~\ref{sec:phases} already provide a way to identify pass-through
transactions, as depicted in Figure~\ref{fig:passthrough}.  In particular, 
running the augmented heuristic for
Phase~1 transactions identifies not only the deposit transaction in the
$\curin$ blockchain but also the Phase~2 transaction (i.e., the withdrawal 
transaction in the $\curout$
blockchain), as this is exactly what is returned by the API.
As discussed above, this has the effect on anonymity of tracing the flow of 
funds across this ShapeShift transaction and linking its two 
endpoints; i.e., the input address(es) in 
the $\curin$ blockchain with the output address(es) in the $\curout$ 
blockchain.  The results, in terms of the percentages of all possible 
transactions between a pair $(\curin,\curout)$ for which we found the
corresponding on-chain transactions, are in Figure~\ref{fig:passthrough-heatmap}.

The figure demonstrates that our success in identifying these types of
transactions varied somewhat, and depended\dash not unsurprisingly\dash on
our success in identifying transactions in the $\curin$ blockchain.  This
means that we were typically least successful with $\curin$ blockchains with
higher transaction volumes, such as Bitcoin, because  
we frequently ended up with multiple
hits (although here we were still able to identify more than 74\% of
transactions).  
In contrast, the dark stripes for Dash and
Zcash demonstrate our high level of success in identifying pass-through 
transactions with those currencies as $\curin$, due to our high level of
success in their Phase~1 analysis in general (89\% and 91\% respectively).  
In total, across all eight currencies we were
able to identify 1,383,666 pass-through transactions. 

\subsection{U-turns}\label{sec:uturn}

As depicted in Figure~\ref{fig:uturn}, we consider a U-turn to be a pattern 
in which a user has just sent money from 
$\curin$ to $\curout$, only to turn around and go immediately back to 
$\curin$.  This means linking two transactions:  the Phase~2 
transaction used to send money to $\curout$ and the Phase~1 transaction
used to send money back to $\curin$.  In terms of timing and amount, we require 
that the second transaction happens within 30 minutes of the first, and that 
it carries within 1\% of the value that was generated by the first Phase~2
transaction. This value is returned by the ShapeShift API in the 
$\outgoingCoin$ field.

While the close timing and amount already give some indication that these two
transactions are linked, it is of course possible that this is a coincidence 
and they were in fact carried out by different users.  In order to gain 
additional confidence that it was the same user, we have two options.
In UTXO-based cryptocurrencies (see Section~\ref{sec:back-currency}), we could
see if the input is the same UTXO that was created in the Phase~2 transaction,
and thus see if a user is spending the coin immediately.
In cryptocurrencies based instead on accounts, such as Ethereum, 
we have no choice but to look just at the addresses.  Here we thus define a
U-turn as seeing if the address that was used as the output in the Phase~2
transaction is used as the input in the later Phase~1 transaction.

Once we identified such candidate pairs of transactions $(\tx_1,\tx_2)$, 
we then ran the augmented heuristic from Section~\ref{sec:phases} to
identify the relevant output address in the $\curout$ blockchain, according
to $\tx_1$.  We then ran the same heuristic to identify the relevant input 
address in the $\curout$ blockchain, this time according to $\tx_2$.

\begin{figure}[t]
	\centering
	\includegraphics[width=0.8\linewidth]{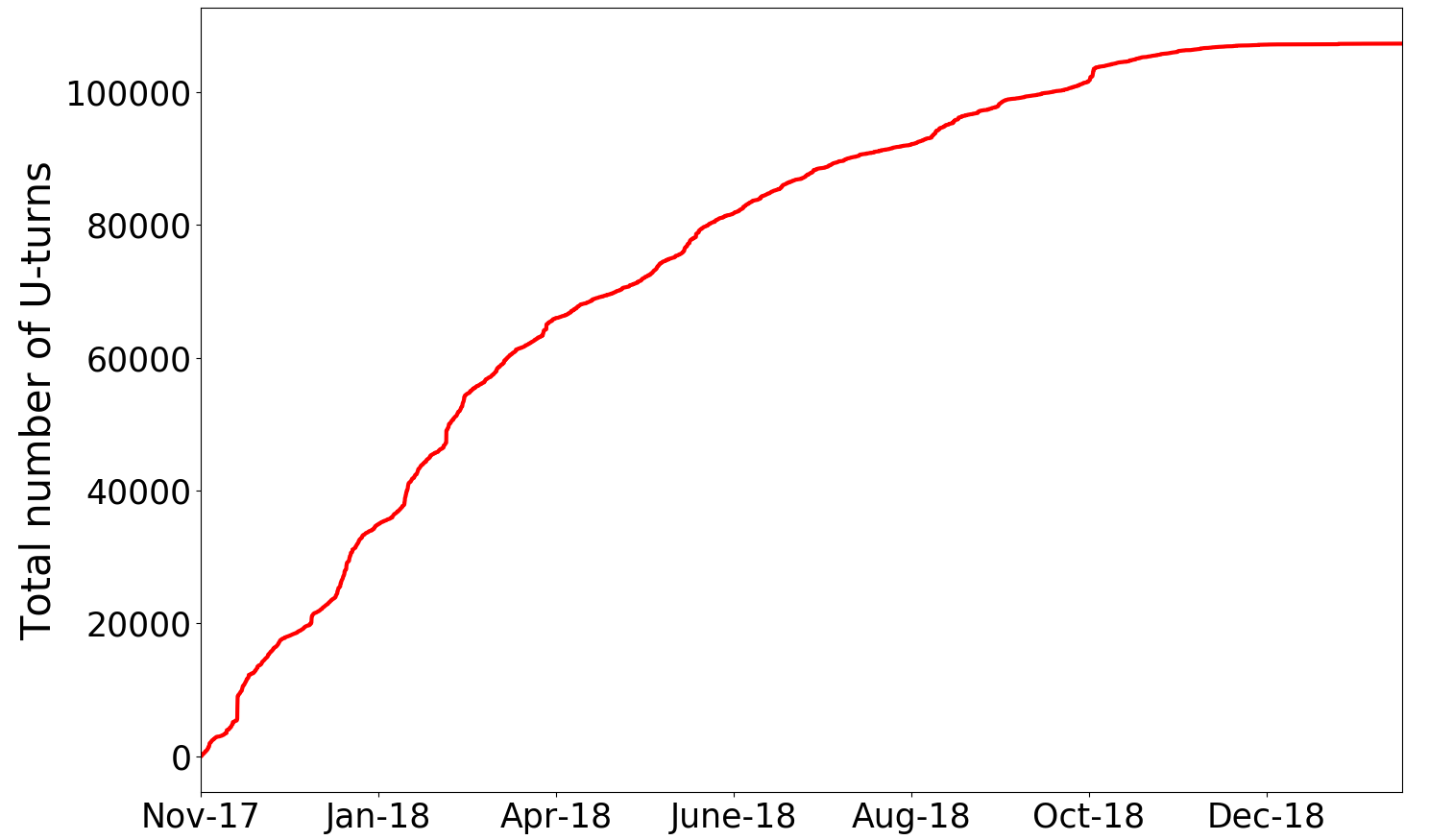}
	\caption{The total number of U-turns over time, as identified by our basic 
		heuristic.}
	\label{fig:AllUturns}
\end{figure}

\begin{figure}[t]
	\centering
	\includegraphics[width=0.8\linewidth]{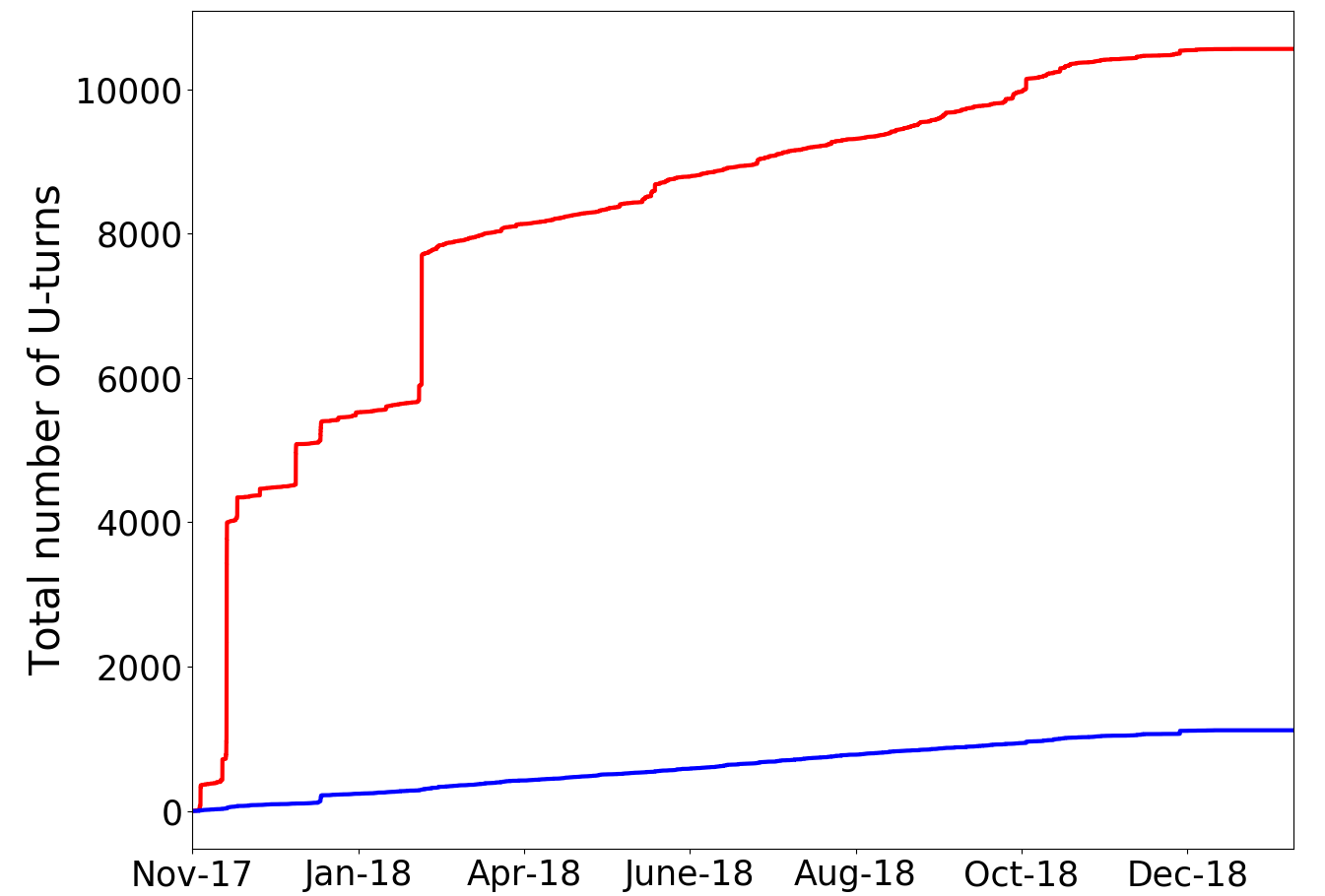}
	\caption{The total number of U-turns over time, as identified by our 
		address-based (in red) and UTXO-based (in blue) heuristics.}
	\label{fig:VerUturns}
\end{figure}

In fact though, what we really identified in Phase~2 was not just an address but, 
as described above, a newly created UTXO.  If the input used in $\tx_2$ was this 
same UTXO, then we found a U-turn according to the first heuristic.  If instead 
it corresponded just to the same address, then we found a U-turn according
to the second heuristic.  The results of both of these heuristics, in addition
to the basic identification of U-turns according to the timing and amount, can 
be found in Table~\ref{tab:UTurnsPerCurrency}, and plots showing their
cumulative number over time can be found in Figures~\ref{fig:AllUturns}
and~\ref{fig:VerUturns}.  In total, we identified 107,267 U-turns according 
to our basic heuristic,
10,566 U-turns according to our address-based heuristic, and 
1,120 U-turns according to our UTXO-based heuristic. 

\begin{table}[t]
	\centering
	\begin{tabular}{lS[table-format=5]S[table-format=3]S[table-format=3]}
		\toprule
		Currency & {\# (basic)} & {\# (addr)} & {\# (utxo)}\\
		\midrule
		BTC & 36666 & 565 & 314 \\
		BCH & 2864 & 196 & 81 \\
		DASH & 3234 & 2091 & 184 \\
		DOGE & 546 & 75 & 75 \\
		ETH & 53518 & 5248 & {-} \\
		ETC & 1397 & 543 & {-} \\
		LTC & 8270 & 1429 & 244 \\
		ZEC & 772 & 419 & 222\\
		\bottomrule
	\end{tabular}
	\caption{The number of U-turns identified for each cryptocurrency, according
		to our basic heuristic concerning timing and value, and both the address-based 
		and UTXO-based heuristics concerning identical ownership.  Since Ethereum and
		Ethereum Classic are account-based, the UTXO heuristic cannot be applied to
		them.}
	\label{tab:UTurnsPerCurrency}
\end{table}

While the dominance of both Bitcoin and Ethereum should be expected given
their overall trading dominance, we also observe that both Dash and Zcash 
have been used extensively as
``mixer coins'' in U-turns, and are in fact more popular for this purpose than
they are overall.
Despite this indication that users may prefer to use privacy coins as the
mixing intermediary, Zcash has the highest percentage of identified UTXO-based 
U-turn transactions.  Thus, these users not only do not gain
extra anonymity by using it, but in fact are easily identifiable given that 
they did not change the address used in 419 out of 772 (54.24\%) cases, 
or\dash even worse\dash immediately shifted back the exact same coin they 
received in 222 (28.75\%) cases.  
In the case of Dash, the results suggest something a bit different.  Once
more, the usage of a privacy coin was not very
successful since in 2091 out of the 3234 cases the address that received the
fresh coins was the same as the one that shifted it back.  It was the exact
same coin in only 184 cases, however, which suggests that although the user is
the same, there is a local Dash transaction between the two ShapeShift
transactions.  We defer a further discussion of this asymmetry to
Section~\ref{sec:zcash-dash}, where we also discuss more generally the use of 
anonymity features in both Zcash and Dash.

Looking at Figure~\ref{fig:VerUturns}, we can see a steep rise in the number
of U-turns that used the same address in December 2017, which is not
true of the ones that used the same UTXO or in the overall number of
U-turns in Figure~\ref{fig:AllUturns}.  Looking into this further, we observed
that the number of U-turns was particularly elevated during this period for
four specific pairs of currencies: DASH-ETH, DASH-LTC, ETH-DASH, and 
LTC-ETH.  This thus affected primarily the address-based heuristic due to the
fact that (1) Ethereum is account-based so the UTXO-based heuristic does not
apply, and (2) Dash has a high percentage of U-turns 
using the same address, but a much smaller percentage using the same UTXO.  
The amount of money shifted in these U-turns varied significantly in terms of
the units of the input currency, but all carried between 115K and 138K in USD.  
Although the ShapeShift transactions that were
involved in these U-turns had hundreds of different addresses in the $\curin$ 
blockchain, they used only a small number of addresses in the $\curout$
blockchain: $4$ addresses in Ethereum, 13 in Dash, and 9 in Litecoin.
As we discuss further in Section~\ref{sec:new-heuristic}, the re-use of 
addresses and the fact that the total amount of money (in
USD) carried by the transactions was roughly the same indicates that perhaps a
small group of people was responsible for creating this spike in the graph.

\subsection{Round-trip transactions}\label{sec:rtt}

As depicted in Figure~\ref{fig:roundtrip}, a round-trip transaction requires
performing two ShapeShift transactions: one out of
the initial currency and one back into it.  
To identify round-trip transactions, we effectively combine the results of the
pass-through and U-turn transactions; i.e., we tagged something as a
round-trip transaction if the output of a pass-through
transaction from X to Y was identified as being involved in a U-turn 
transaction, which was itself linked to a later pass-through transaction from
Y to X (of roughly the same amount).  As described at the beginning of the
section, this has the powerful effect of creating a link between the sender
and recipient within a single currency, despite the fact that money flowed
into a different currency in between.

In more detail, we looked for consecutive ShapeShift transactions 
where for a given pair of cryptocurrencies X and Y: 
(1) the first transaction was of the form X-Y; (2) the second transaction 
was of the form Y-X; (3) the second transaction happened relatively
soon after the first one; and (4) the value carried by the two 
transaction was approximately the same.  For the third property, we required
that the second transaction happened within 30 minutes of the first.  For the
fourth property, we
required that if the first transaction carried $x$ units of $\curin$ then the
second transaction carried within $0.5\%$ of the value in the (on-chain) 
Phase~2 transaction, according to the $\outgoingCoin$ field provided by the
API.  

As with U-turns, we considered an additional restriction to capture the case
in which the user in the $\curin$ blockchain stayed the same, meaning money 
clearly did not change hands.  Unlike with U-turns, however, this restriction 
is less to provide accuracy for the basic heuristic and more to isolate the 
behavior of people engaged in day trading or money laundering (as opposed to
those meaningfully sending money to other users).  For 
this pattern, we identify the input addresses used in Phase~1 for the first
transaction, which represent the user who initiated the round-trip transaction 
in the $\curin$ blockchain.  We then identify the output addresses used in
Phase~2 for the second transaction, which represent the user who was the
final recipient of the funds.  If the address was the same, then it is clear
that money has not changed hands.  Otherwise, the round-trip transaction acts
as a heuristic for linking together the input and output addresses.

\begin{table}[t]
	\centering
	\begin{tabular}{lS[table-format=5]S[table-format=2]S[table-format=2]}
		\toprule
		Currency & {\# (regular)} & {\# (same addr)} \\
		\midrule
		BTC & 35019 & 437  \\
		BCH & 1780 & 84  \\
		DASH & 3253 & 2353  \\
		DOGE & 378 & 0 \\
		ETH & 45611 & 4085 \\
		ETC & 1122 & 626  \\
		LTC & 6912 & 2733  \\
		ZEC & 472 & 172 \\
		\bottomrule
	\end{tabular}
	\caption{The number of regular round-trip transactions identified for each 
		cryptocurrency, and the number that use the same initial and final address.}
	\label{tab:RTTsPerCurrency}
\end{table}

The results of running this heuristic (with and without the extra restriction) 
are in Table~\ref{tab:RTTsPerCurrency}.  In total, we identified 95,547
round-trip transactions according to our regular heuristic, and identified 
10,490 transactions where
the input and output addresses were the same.  Across different currencies,
however, there was a high level of variance in the results.  While this could 
be a result of the different levels of accuracy in Phase~1 for 
different currencies, the 
more likely explanation is that users indeed engage in different patterns of
behavior with different currencies.  For Bitcoin, for example, there was a 
very small percentage (1.2\%) of round-trip transactions that used the
same address.  This suggests that either users are aware of the
general lack of anonymity in the basic Bitcoin protocol and use ShapeShift to
make anonymous payments, or that if they do use round-trip transactions as a 
form of money laundering they are at least careful enough to change their
addresses.  More simply, it may just be the case that generating new addresses
is more of a default in Bitcoin than it is in other currencies.

In other currencies, however, such as Dash, Ethereum Classic, Litecoin, and
Zcash, there were relatively high percentages of round-trip transactions that 
used the same 
input and output address: 72\%, 56\%, 40\%, and 36\% respectively.  In Ethereum
Classic, this may be explained by the account-based nature of the currency, 
which means that it is common for one entity to use only one address, although 
the percentage for Ethereum is much lower (9\%).  In Dash and Zcash, as we 
have already seen in Section~\ref{sec:uturn} and explore further in
Section~\ref{sec:zcash-dash}, it may simply be the case that users assume they
achieve anonymity just through the use of a privacy coin, so do not take extra
measures to hide their identity.

\section{Clustering Analysis}\label{sec:clusters}

\subsection{Shared ownership heuristic}\label{sec:multi-input}

As described in Sections~\ref{sec:changelly-data}
and~\ref{sec:shapeshift-data}, we engaged in transactions with both ShapeShift 
and Changelly, which provided us with some ground-truth evidence of addresses
that were owned by them.  We also collected three sets of
tagging data (i.e., tags associated with addresses that describe their
real-world owner): for Bitcoin we used the data available from
WalletExplorer,\footnote{\url{https://www.walletexplorer.com/}} which covers a
wide variety of different Bitcoin-based services; for Zcash we used
hand-collected data from Kappos et al.~\cite{sarah-zcash}, which covers only
exchanges; and for Ethereum we used the data available from
Etherscan,\footnote{\url{https://etherscan.io/}} which covers a variety of
services and contracts.

In order to understand the behavior of these trading platforms and the
interaction they had with other blockchain-based services such as exchanges,
our first instinct was to combine these tags with the now-standard 
``multi-input'' clustering heuristic for 
cryptocurrencies~\cite{reid2013analysis,meiklejohn-fistful}, which states that 
in a transaction with multiple input addresses, all inputs belong to the same
entity.  %
When we applied this clustering heuristic to an earlier version of our 
dataset~\cite{anon-us}, however, the results were fairly uneven.  For
Dogecoin, for example, the three ShapeShift transactions we performed revealed 
only three addresses, which each had done a very small number of
transactions.  The three Changelly transactions we performed, in contrast,
revealed 24,893 addresses, which in total had received over 67
trillion DOGE.  These results suggest that the trading platforms operate a
number of different clusters in each cryptocurrency, and perhaps even change
their behavior depending on the currency, which in turns makes it clear that
we did not capture a comprehensive view of the activity of either.

More worrying, in one of our Changelly transactions, we received coins from a 
Ethereum address that had been tagged as belonging to HitBTC, a prominent exchange.
This suggests that Changelly may occasionally operate using exchange accounts,
which would completely invalidate the results of the clustering heuristic, as
their individually operated addresses would end up in the same cluster as all
of the ones operated by HitBTC.  We thus decided not to use this type of
clustering, and to instead focus on a new clustering heuristic geared at
identifying common social relationships.

\subsection{Common relationship heuristic}\label{sec:new-heuristic}

As it was clear that the multi-input heuristic would not yield meaningful
information about shared ownership, we chose to switch our focus away from the
interactions ShapeShift had on the blockchain and look instead at
the relationships between individual ShapeShift users.  In particular, we
defined the following heuristic:

\begin{heuristic}\label{heuristic:clusters}
	If two or more addresses send coins to the same address in the $\curout$
	blockchain, or if two or more addresses receive coins from the same address 
	in the $\curin$ blockchain, then these addresses have some common social 
	relationship.
\end{heuristic} 

The definition of a common social relationship is (intentionally) vague, and 
the implications of this heuristic are indeed less clear-cut than those of
heuristics around shared ownership.
Nevertheless, we consider what it means for two different addresses, in 
potentially two different blockchains, to have sent coins to the same address;
we refer to these addresses as belonging in the \emph{input} cluster of the
output address (and analogously refer to the \emph{output} cluster for an
address sending to multiple other addresses).
In the case in which the addresses are most closely linked, it could represent 
the same user consolidating money held across different currencies into a 
single one.  It could also represent different users interacting with a common 
service, such as an exchange.  Finally, it could simply be two users who do 
not know each other directly but happen to be sending money 
to the same individual.  What cannot be the case, however, is that the
addresses are not related in any way.

To implement this heuristic, we parsed transactions into a graph where we
defined a node as an address and a directed edge $(u,v)$ as existing when 
one address $u$ initiated a ShapeShift transaction sending coins to $v$, which
we identified using the results of our pass-through analysis from 
Section~\ref{sec:phases}. (This means that the inputs in our graph are
restricted to those for which we ran Phase~1 to find the address, and thus 
that our input clusters contain only the top 8 currencies.  
In the other direction, however, we obtain the address directly from the API, 
which means output clusters can contain all currencies.)
Edges are further weighted by the number of transactions sent from $u$ to $v$.  
For each node, the cluster centered on that address was then defined as all
nodes adjacent to it (i.e., pointing towards it).

Performing this clustering generated a graph with 2,895,445
nodes (distinct addresses) and 2,244,459 edges.  
Sorting the clusters by in-degree reveals the entities that received
the highest number of ShapeShift transactions (from the top 8
currencies, per our caveat above).
The largest cluster had 12,868 addresses\dash many of them belonging to
Ethereum, Litecoin, and Dash\dash and was centered on a Bitcoin 
address belonging to CoinPayments.net, a multi-coin payment processing
gateway.  Of the ten largest clusters, three others (one associated with 
Ripple and two with Bitcoin addresses) are also connected with CoinPayments, 
which suggests that ShapeShift is a popular platform amongst its users.  

Sorting the individual clusters by out-degree reveals instead the users who 
initiated the highest number of ShapeShift transactions.
Here the largest cluster (consisting of 2314 addresses) was centered on a 
Litecoin address, and the second largest cluster was centered on an Ethereum 
address that belonged to Binance (a popular exchange).  Of the
ten largest clusters, two others were centered on Binance-tagged addresses,
and three were centered on other exchanges (Freewallet, Gemini, and Bittrex).
While it makes sense that exchanges typically dominate on-chain activity in
many cryptocurrencies, it is somewhat surprising to also observe that
dominance here, given that these exchanges already allow users to shift between 
many different cryptocurrencies.  Aside from the potential for better rates or
the perception of increased anonymity, it is thus unclear why a user wanting to
shift from one currency to another would do so using ShapeShift
as opposed to using the same service with which they have already stored 
their coins.

Beyond these basic statistics, we apply this heuristic to several of the 
case studies we investigate in the next section.  We also revisit here
the large spike in the number of U-turns that we observed in
Section~\ref{sec:uturn}.  Our hypothesis then was that this spike was caused
by a small number of parties, due to the similar USD value carried by the
transactions and by the re-use of a small number of addresses across Dash,
Ethereum, and Litecoin.  Here we briefly investigate this further by 
examining the clusters centered on these addresses.  

Of the 13 Dash addresses, 
all but one of them formed small input and output clusters that were comprised 
of addresses solely from Litecoin and Ethereum.  
Of the 9 Litecoin addresses, 6 had input clusters consisting solely
of Dash and Ethereum addresses, with two of them consisting solely of Dash 
addresses.  
Finally, of the 4
Ethereum addresses, all of them had input clusters consisting solely of Dash
and Litecoin addresses.  One of them, however, had a diverse set of addresses 
in its output cluster, belonging to Bitcoin, Bitcoin Cash, and a number of 
Ethereum-based tokens.  These results thus still suggest a small number of 
parties, due to the tight connection between the three currencies in the 
clusters, although of course further investigation would be needed to get a
more complete picture.

\section{Patterns of ShapeShift Usage}\label{sec:case-studies}

In this section, we examine potential applications of the analysis developed
in previous sections, in terms of identifying specific usages of ShapeShift.
As before, our focus is on anonymity, and the potential that such platforms
may offer for money laundering or other illicit purposes, as well as for
trading.  To this end, we
begin by looking at two case studies associated with explicitly criminal
activity and examine the interactions these criminals had with the ShapeShift
platform.  We then switch in Section~\ref{sec:trading-bots} to look at
non-criminal activity, by attempting to identify trading bots that use
ShapeShift and the patterns they may create.  Finally, in
Section~\ref{sec:zcash-dash} we look at the role
that privacy coins (Monero, Zcash, and Dash) play, in order to 
identify the extent to
which the usage of these coins in ShapeShift is motivated by a desire for
anonymity.

\subsection{Starscape Capital}

In January 2018, an investment firm called Starscape Capital raised over
2,000~ETH (worth 2.2M~USD at the time) during their Initial Coin Offering,
after promising users a 50\% return in exchange for investing in their
cryptocurrency arbitrage fund.  Shortly afterwards, all of their social media 
accounts disappeared, and it was reported that an amount of ETH worth
517,000~USD was sent from their wallet to ShapeShift, where it was shifted into 
Monero~\cite{scheck_shifflett_2018}.  

We confirmed this for ourselves by observing that the address known to be 
owned by Starscape Capital participated in 192 Ethereum transactions across a 
three-day span (January 19-21), during which it received and sent 
2,038 ETH; in total it sent money in 133 transactions.  We found that 109 
of these transactions sent money to 
ShapeShift, and of these 103 were shifts to Monero conducted on January 21 
(the remaining 6 were shifts to Ethereum). The total amount shifted into 
Monero was 465.61~ETH (1388.39~XMR), and all
of the money was shifted into only three different Monero addresses, of which
one received 70\% of the resulting XMR.  Using the clusters defined in 
Section~\ref{sec:new-heuristic}, we did not find evidence of any other 
addresses (in any other currencies)
interacting with either the ETH or XMR addresses associated with Starscape
Capital.  %

\subsection{Ethereum-based scams}

EtherScamDB\footnote{https://etherscamdb.info/} is a website that, based on 
user reports that are manually investigated by its operators, collects and 
lists Ethereum addresses that have been involved in scams.
As of January 30 2019, they had a total
of 6374 scams listed, with 1973 associated addresses.  
We found that 194 of these addresses (9\% of those listed) had been involved 
in 853 transactions to ShapeShift, of which 688 had a $\status$ field of 
\verb#complete#.
Across these successful transactions, 1797~ETH was shifted to other
currencies: 74\% to Bitcoin, 19\% to Monero, 3\% to Bitcoin Cash,
and 1\% to Zcash.

The scams which successfully shifted the highest volumes belonged to so-called 
trust-trading and MyEtherWallet scams. Trust-trading is a scam based on
the premise that users who send coins prove the legitimacy of their
addresses, after which the traders ``trust'' their address and send 
back higher amounts (whereas in fact most users send money and simply receive
nothing in return).  
This type of scam shifted over 918~ETH, the majority of which was converted 
to Bitcoin (691~ETH, or 75\%). 
A MyEtherWallet scam is a phishing/typosquatting scam where scammers
operate a service with a similar name to the popular online wallet
MyEtherWallet,\footnote{https://www.myetherwallet.com/} in order to trick
users into giving them their account details.
These scammers shifted the majority of the stolen ETH to Bitcoin
(207~ETH) and to Monero (151~ETH).

Given that the majority of the overall stolen coins was shifted to Bitcoin, we 
next investigated whether or not these stolen coins could be tracked further
using our analysis.  %
In particular, we looked to see if they performed a U-turn or a round-trip
transaction, as discussed in Section~\ref{sec:tracking}. We identified one 
address, associated with a trust-trading scam, that participated in 34 
distinct round-trip transactions, all coming back to a different address from
the original one.  All
these transactions used Bitcoin as $\curout$ and used the same address in
Bitcoin to both receive and send coins; i.e., we identified the U-turns in
Bitcoin according to our address-based heuristic.  In total, more than 70~ETH
were circulated across these round-trip transactions.

\subsection{Trading bots}\label{sec:trading-bots}

ShapeShift, like any other cryptocurrency exchange, can be used by 
traders who wish to take advantage of the volatility in cryptocurrency prices.
The potential advantages of doing this via ShapeShift, 
as compared with other platforms that focus more on the exchange between
cryptocurrencies and fiat currencies, are that (1) ShapeShift transactions can
be easily automated via their API, and (2) a single ShapeShift transaction
acts to both purchase desired coins and dump unwanted ones.  
Such trading usually requires large volumes of transactions and high precision 
on their the timing, due to the constant fluctuation in cryptocurrency prices.  
We thus looked for activity that involved large numbers of similar
transactions in a small time period, on the theory that it would be
associated primarily with trading bots.

We started by searching for sets of consecutive ShapeShift
transactions that carried approximately the same value in $\curin$ (with an error
rate of 1\%) and involved the same currencies.  When we did this, however, we 
found thousands of such sets.  We thus added the extra conditions that there
must be at least 15 transactions in the set that took place in a
span of five minutes; i.e., that within a five-minute block of ShapeShift
transactions there were at least 15 involving the same currencies and carrying
the same approximate USD value.  This resulted in 107 such sets.  

After obtaining our 107 trading clusters, we removed transactions that 
we believed were false positives in that they happened to have a similar 
value but were clearly the odd one out.  For example, in a cluster of 20 
transactions with 19 ETH-BTC transactions and one LTC-ZEC transaction, 
we removed the latter.  We were thus left with clusters of either a 
particular pair (e.g., ETH-BTC) or two pairs where the $\curout$ or the 
$\curin$ was the same (e.g., ETH-BTC and ZEC-BTC), which suggests either 
the purchase of a rising coin or the dump of a declining one.  We sought to
further validate these clusters by using our heuristic from
Section~\ref{sec:new-heuristic} to see if the clusters shared common
addresses.  While we typically did not find this in UTXO-based currencies (as
most entities operate using many addresses), in account-based currencies 
we found that in almost every case there was one particular address that was 
involved in the trading cluster. 

\begin{figure}[t]
	\centering
	\includegraphics[width=\linewidth]{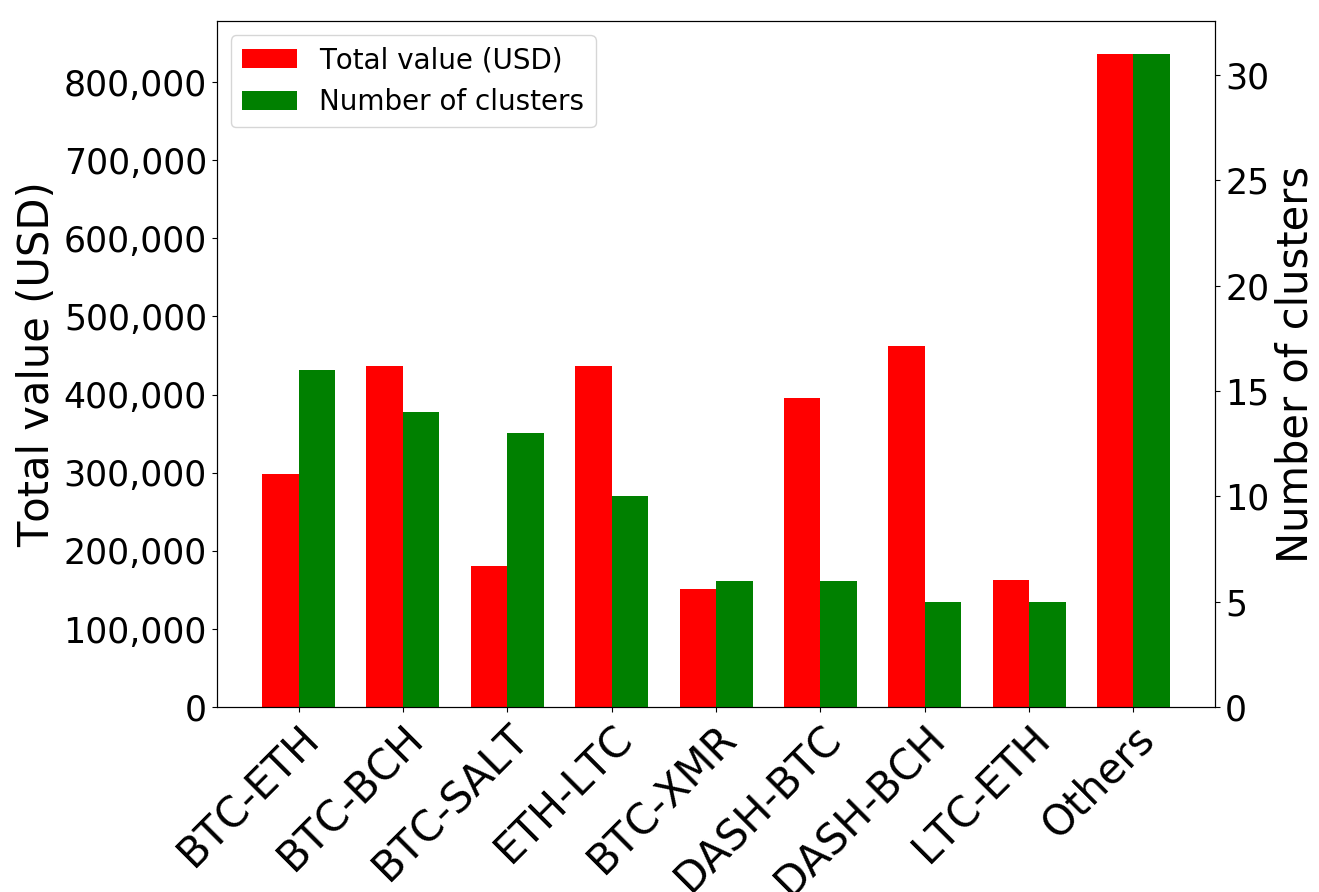}
	\caption{Our 107 clusters of likely trading bots, categorized by the
		pair of currencies they trade between and the total amount transacted by 
		those clusters (in USD).}
	\label{fig:barchart-bots}
\end{figure}

We summarize our results in Figure~\ref{fig:barchart-bots}, in terms of the 
most common pairs of currencies and the total money exchanged by trading 
clusters using those currencies.  It is clear that 
the most common interactions are performed between the most popular currencies
overall, with the exception of Monero (XMR) and SALT.  In particular, we found 
six clusters consisting of 17-20 transactions that exchanged BTC 
for XMR, and 13 clusters that exchanged BTC for SALT, an Ethereum-based 
token.
The sizes of each trading cluster varied between 
16 and 33 transactions and in total comprise 258 transactions, each of which
shifted exactly 0.1~BTC.  In total they originated from 514 
different Bitcoin addresses, which may make it appear as though 
different people carried out these transactions.  After applying our 
pass-through heuristic, however, we
found that across all the transactions there were only two distinct SALT
addresses used to receive the output.  It is thus instead likely that this
represents trading activity involving one or two entities.
%
%
%
%
%
%
%

\subsection{Usage of anonymity tools}\label{sec:zcash-dash}

Given the potential usage of ShapeShift for money laundering or other criminal
activities, we sought to understand the extent to which its users seemed
motivated to hide the source of their funds.  While using ShapeShift is
already one attempt at doing this, we focus here on the combination of using
ShapeShift and so-called ``privacy coins'' (Dash, Monero, and Zcash) that are 
designed to offer improved anonymity guarantees.  

In terms of the effect of the introduction of KYC into ShapeShift, the
number of transactions using Zcash as $\curin$ averaged 164 per day 
the month before, and averaged 116 per day the month after.  We also saw a 
small decline with Zcash as $\curout$: 69 per day before and 43 per day 
after.  Monero and Dash, however, saw much higher declines, and
in fact saw the largest declines across all eight cryptocurrencies.  The 
daily average the month before was $136$ using Monero as $\curin$, whereas it 
was $47$ after.  Similarly, the daily average using it as $\curout$ was 
$316$ before and $62$ after. For Dash, the daily average as $\curin$ 
was $128$ before and $81$ after, and the daily average as $\curout$ 
was $103$ before and $42$ after.

In terms of the blockchain data we had (according to the most popular 
currencies), our analysis in what follows is restricted to Dash and Zcash,
although we leave an exploration of Monero as interesting future work.

\subsubsection{Zcash}

The main anonymity feature in Zcash is known as the \emph{shielded pool}.
Briefly, transparent Zcash transactions behave just like Bitcoin transactions
in that they reveal in the clear the sender and recipient (according to
so-called \emph{t-addresses}), as well as the value being sent.  This
information is hidden to various degrees, however, when interacting with
the pool.  In particular, when putting money into the pool the recipient is
specified using a so-called \emph{z-address}, which hides the recipient but still
reveals the sender, and taking money out of the pool hides the sender (through
the use of zero-knowledge proofs~\cite{SP:BCGGMT14}) but reveals the
recipient.  Finally, Zcash is designed to provide privacy mainly in the case
in which users transact \emph{within} the shielded pool, which hides the
sender, recipient, and the value being sent.

\begin{figure}
	\centering
	\includegraphics[width=0.7\linewidth]{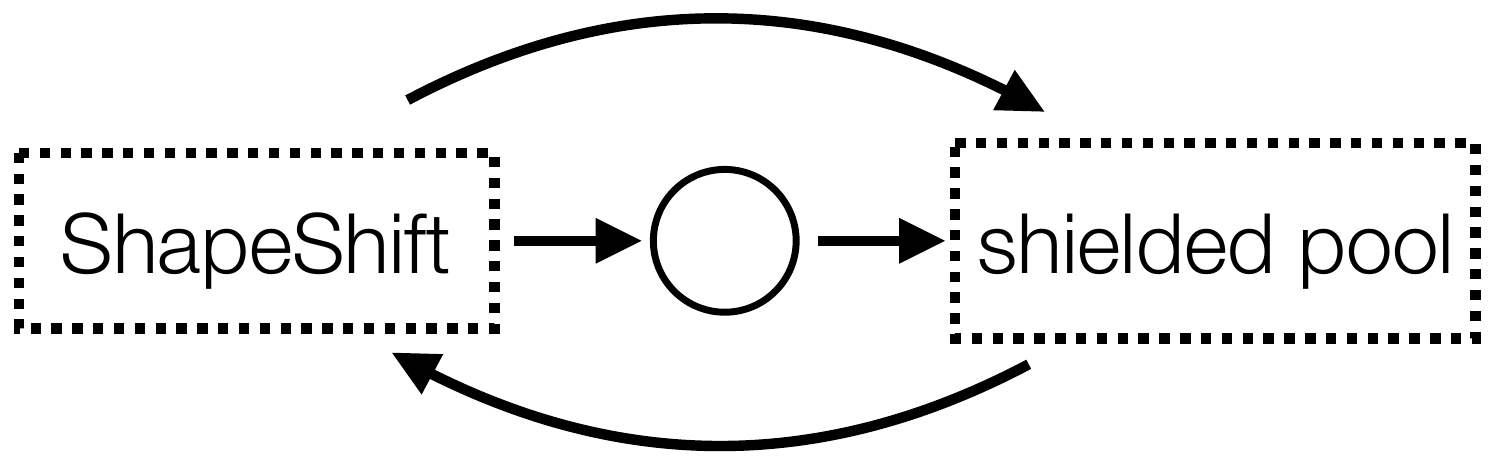}
	\caption{The three types of interactions we investigated between ShapeShift and
		the shielded pool in Zcash.}
	\label{fig:nugget}
\end{figure}

We considered three possible interactions between ShapeShift and the shielded
pool, as depicted in
Figure~\ref{fig:nugget}: (1) a user shifts coins directly from ShapeShift into
the shielded pool, (2) a user shifts to a t-address but then uses that
t-address to put money into the pool, and (3) a user sends money directly from
the pool to ShapeShift.

For the first type of interaction, we found 29,003 transactions that used ZEC
as $\curout$.  Of these, 758 had a z-address as the output address, meaning
coins were sent directly to the shielded pool.  The total value put into the
pool in these transactions was 6,707.86~ZEC, which is 4.3\% of all the ZEC
received in pass-through transactions.  When attempting to use
z-addresses in our own interactions with ShapeShift, however, we encountered
errors or were told to contact customer service.  It is thus not clear if
usage of this feature is supported at the time of writing.

For the second type of interaction, 
there were 1309 where the next transaction (i.e., the transaction in which this
UTXO spent its contents) involved putting money into the pool.  The total
value put into the pool in these transactions was 12,534~ZEC, which is
8.2\% of all the ZEC received in pass-through transactions.

For the third type of interaction, we found 111,041 pass-through transactions
that used ZEC as $\curin$.  Of these, 3808 came directly
from the pool, with a total value of 22,490~ZEC (14\% of all the
ZEC sent in pass-through transactions).

Thus, while the usage of the anonymity features in Zcash was not
necessarily a large fraction of the overall usage of Zcash in ShapeShift,
there is clear potential to move large amounts of Zcash (representing over 10
million USD at the time it was transacted) by combining ShapeShift with the
shielded pool.  

\subsubsection{Dash}

As in Zcash, the ``standard'' transaction in Dash is similar to a Bitcoin
transaction in terms of the information it reveals.  Its main anonymity
feature\dash \emph{PrivateSend} transactions\dash are a type of 
CoinJoin~\cite{maxwell2013coinjoin}.  A CoinJoin is specifically designed to
invalidate the multi-input clustering heuristic described in
Section~\ref{sec:clusters}, as it allows multiple users to come together and
send coins to different sets of recipients in a single transaction.  If each
sender sends the same number of coins to their recipient, then it is difficult
to determine which input address corresponds to which output address, thus
severing the link between an individual sender and recipient.

In a traditional CoinJoin, users must find each other in some offline manner
(e.g., an IRC channel) and form the transaction together over several rounds 
of communication.  
This can be a cumbersome process, so Dash aims to simplify it for users by
automatically finding other users for them and chaining multiple mixes
together.  In order to ensure that users cannot accidentally de-anonymize
themselves by sending uniquely identifiable values, these PrivateSend
transactions are restricted to specific denominations: 0.01, 0.1, 1, and
10~DASH.  As observed by Kalodner et al.~\cite{blocksci}, however, the CoinJoin 
denominations often contain a fee of 0.0000001 DASH, which must be factored in 
when searching for these transactions.  Our parameters for identifying a 
CoinJoin were thus that (1) the transaction must have at least three inputs, 
(2) the outputs must consist solely of values from the list of possible 
denominations (modulo the fees), and (3) and all output values must be 
the same.  
In fact, given how Dash operates there is always one output with a
non-standard value, so it was further necessary to relax the second and third
requirements to allow there to be at most one address that does not carry the
specified value.

We first looked to see how often the DASH sent to ShapeShift had 
originated from a CoinJoin, which meant identifying if the inputs of a 
Phase~1 transaction were outputs from a CoinJoin.  
Out of 100,410 candidate transactions, we found 2,068 that came from a 
CoinJoin, carrying a total of 11,929~DASH in value (6.5\% of the total value
across transactions with Dash as $\curin$).
Next, we looked at whether or not users performed a CoinJoin after 
receiving coins from ShapeShift, which meant identifying if the outputs of a
Phase~2 transaction had been spent in a CoinJoin.  
Out of 50,545 candidate 
transactions, we found only 33 CoinJoin transactions, carrying a total of
187~DASH in value (0.1\% of the total value across transactions using Dash as
$\curout$).

If we revisit our results concerning the use of U-turns in Dash 
from Section~\ref{sec:uturn}, we recall that there was a large asymmetry in
terms of the results of our two heuristics: only 5.6\% of the U-turns used the
same UTXO, but 64.6\% of U-turns used the same address.  This suggests that some
additional on-chain transaction took place between the two ShapeShift 
transactions, and indeed upon further inspection we identified many cases
where this transaction was a CoinJoin.  There thus appears to have
been a genuine attempt to take advantage of the privacy that Dash offers, but
this was completely ineffective due to the use of the same address that both 
sent and received the mixed coins.

\section{Conclusions}\label{sec:conclusions}

In this study, we presented a characterization of the usage of the ShapeShift
trading platform over a thirteen-month period, focusing on the ability to link
together the ledgers of multiple different
cryptocurrencies.  To accomplish this task, we looked at these trading 
platforms from several different perspectives, ranging from the correlations 
between the transactions they produce in the cryptocurrency ledgers to the 
relationships they reveal between seemingly distinct users.  The 
techniques we develop demonstrate that it is possible to
capture complex transactional behaviors and trace their activity even as it 
moves across ledgers, which has implications for any
criminals attempting to use these platforms to obscure their flow of money.

\section*{Acknowledgments}

We would like to thank Bernhard Haslhofer and Rainer St{\"u}tz for performing
the Bitcoin multi-input clustering using the GraphSense tool, and Zooko Wilcox, 
the anonymous reviewers, and our shepherd Matthew Green for their feedback.  
All authors are supported by the EU H2020 TITANIUM project under grant 
agreement number 740558. 

{
\balance
{\footnotesize
\bibliographystyle{abbrv}
\begin{flushleft}

\end{flushleft}
}
}

\end{document}